\documentclass[a4p aper,fleqn,usenatbib]{mnras}
\usepackage[T1]{fontenc}
\usepackage{ae,aecompl}
\usepackage{latexsym,amssymb}
\usepackage{graphicx}
\usepackage{ulem}
\usepackage{xcolor}

\newcommand{\msun}{\mbox{$M_\odot$}}
\newcommand{\rsun}{\mbox{$R_\odot$}}
\title[Jets in CE: a low mass MS star in a RG]{Jets in Common Envelopes: a low mass main sequence star in a red giant}
\author[L\'opez-C\'amara et al.]{Diego L\'opez-C\'amara$^{1}$\thanks{E-mail: diego@astro.unam.mx}, Fabio De Colle$^{2}$, Enrique Moreno M\'endez$^{3,4}$, \newauthor{Sagiv Shiber$^{5}$, Roberto Iaconi$^{6}$} \\
$^{1}$C\'atedras CONACyT -- Universidad Nacional Aut\'onoma de M\'exico, Instituto de Astronom\'ia, AP 70-264, CDMX  04510, M\'exico\\
$^{2}$Instituto de Ciencias Nucleares, Universidad Nacional Aut\'onoma de M\'exico, A. P. 70-543 04510 D. F. M\'exico\\
$^{3}$Facultad de Ciencias, Universidad Nacional Aut{\'o}noma de M{\'e}xico, A. P. 70-543,  CDMX 04510,  M\'exico\\
$^{4}$CCyT, UACM, Casa Libertad, Ermita Iztapalapa 4163, Lomas de Zaragoza, Iztapalapa, 09620, CDMX, M\'exico\\
$^{5}$Department of Physics and Astronomy, Louisiana State University, Baton Rouge, LA, 70803 USA\\
$^{6}$Department of Astronomy, Kyoto University, Kitashirakawa-Oiwake-cho, Sakyo-ku, Kyoto 606-8502, Japan}

\date{Accepted XXX. Received YYY; in original form ZZZ}

\pubyear{2020}

\begin{document}
\label{firstpage}
\pagerange{\pageref{firstpage}--\pageref{lastpage}}
\maketitle

\begin{abstract}
We present small-scale three-dimensional hydrodynamical simulations of the evolution of a 0.3 $M_\odot$ main sequence star which launches two perpendicular jets within the envelope of a 0.88 $M_\odot$ red giant. Based on previous large-scale simulations, we study the dynamics of the jets either when the secondary star is grazing, when it has plunged-in, or when it is well-within the envelope of the red giant (in each stage for $\sim$11~days). The dynamics of the jets through the common envelope (CE) depend on the conditions of the environment as well as on their powering. In the grazing stage and the commencement of the plunge self-regulated jets need higher efficiencies to break out of the envelope of the RG. Deep inside the CE, on the timescales simulated, jets are choked independently of whether they are self-regulated or constantly powered. Jets able to break out of the envelope of the RG in large-scale simulations, are choked in our small-scale simulations. The accreted angular momentum onto the secondary star is not large enough to form a disk. The mass accretion onto the MS star is 1-10\% of the Bondi-Hoyle-Littleton rate ($\sim 10^{-3}-10^{-1}$ M$_\odot$~yr$^{-1}$). High luminosity emission, from X-rays to UV and optical, is expected if the jets break out of the CE. Our simulations illustrate the need for inclusion of more realistic accretion and jet models in the dynamical evolution of the CEs.

\end{abstract}


\section{Introduction}
\label{sec:int}

The common envelope (CE) phase is a key stage during the evolution of a binary system \citep{ivanova2013}. It is a short-lived phase in which the envelope of the more massive stellar component evolves into a Red Giant (RG) star, fills its Roche-lobe, and engulfs the less massive star \citep{paczynski1976, ibenlivio1993}. The less massive star, which may be a low mass main sequence (MS) star, inspirals around the core of the most massive, reaches a small orbital separation and may produce a merger, or lead to the ejection of the CE and the emergence of a close binary system \citep{demarcoizzard2017}. Thus, the CE is key in producing high-energy and transitory phenomena (HEAP) such as type Ia supernovae \citep[][]{chevalier2012, py2014}, short GRBs \citep[][]{vigna2020}, long GRBs \citep[][]{brown2007,2011MM}, and gravitational waves \citep[][]{abbott2016}, among others.

The CE evolution is dictated by the release of orbital energy as the secondary orbits within the CE \citep[the so-called ``$\alpha$ formalism'',][]{heuvel76} and by other energy sources \citep{ivanova2013}. Ionization energy \citep{han1995, reichardt2020}, long period pulsations \citep{clayton2017}, dust formation \citep{glanzperets2018, iaconi2020}, accretion onto the secondary star \citep{chamandy2018}, and the launching of jets \citep{soker2004} may have an important role in the CE evolution.

Angular momentum ($J$) is an important ingredient in stellar evolution. In binary systems, the mass transferred from one star to another carries $J$ and thus forms an accretion disk (around the accreting stellar component), from which jets may be launched due to magnetic processes such as the \citet{bp82} mechanism. Jets may shape planetary nebulae \citep[see][and references therein]{soker2020}, and bipolar nebula may be produced \citep{kaminski2021}. In isolated stars jets may play an important role in the explosions of core collapse supernovae \citep{khokhlov1999, papishsoker2011, papishsoker2014, gilkissoker2014, gilkissoker2016, bear2017, piran2017}, or may produce the ears observed in some core collapse supernova remnants \citep{castelletti2006, gonzalez14, grichener2017}, to mention a few examples. 

If the jets are launched when the secondary is in the outskirts of the RG, then the jets will graze the envelope of the more massive progenitor. The jets may remove the envelope mass efficiently so that no CE would be produced and a grazing envelope evolution would take place instead \citep[GEE, see][]{soker2015, shiber2017}. If the jets are launched within the envelope of the RG, they may have an important role in the evolution of the CE phase \citep{armitage2000, soker2004, chevalier2012, shiber2016, soker2018, gilkis2019, schreier2019, soker2019, jones2020}. The Boomerang Nebula, for example, may have been produced by a CE event in which jets played an important role \citep{Sahai2017}.

Previous CE hydrodynamical (HD) simulations have mainly followed the large scale evolution of the system, i.e., including the entire CE in the computational domain  \citep[e.g.,][]{passy2012, kuruwita2016, iaconi2017, iaconi2019, reichardt2019}. Large scale numerical simulation studies allow an understanding of the overall behavior of the CE phase, the loss of material from the system and, eventually, the end of the CE phase. Three-dimensional (3D) HD large scale studies have encountered difficulties in unbinding large percentages of the stellar envelope and thus to fully understand the end of the CE phase \citep{rickertamm2012, ohlmann2016, staff2016, chamandy2018, chamandy2020, prust2019, sand2020, glanzperets2021a, glanzperets2021b}. Also, large scale CE studies involving the presence of jets show that a fraction of the envelope may be blown away during the grazing envelope evolution \citep{shiber2016, shiber2018}, or when the MS star is immersed within the envelope of the massive star \citep{shiber2019, schreier2019, Zou2022}.

Small scale simulations, i.e. simulations with a computational box covering only a fraction of the CE, are useful to understand the details of the propagation of the jets, energy deposition, turbulence, accretion onto the secondary, and the disk formation \citep[e.g.,][]{macleod2015, macleod2017, chamandy2019, de2020, everson2020}. 3D HD small scale simulations of the effects of jets in the CE phase have shown that accretion can be hypercritical and that the jets may play a key role in the evolution of the system \citep{mm2017, lc2019, lc2020}. However, small scale simulations have the disadvantage that only small time scales and the local behavior of the system are followed ($~sim$ a few days, i.e. 10\% of the time needed for the companion star to plunge in the core of the primary; and $\sim 20\%$ of the volume of the red giant in the simulations presented in this paper, respectively).

In this paper we study, by a set of small scale high-resolution 3D HD simulations, the propagation of jets launched from a MS star in the CE of a RG. In this paper we zoom in one part of a simulation of \citet[][S19 from now on]{shiber2019} and study in detail the evolution of the jets and their surroundings by taking advantage of the higher resolution as well as including self-regulation in the jets (which has not been included in large scale CE simulations). As initial conditions of the simulations, we use the CE large scale configuration computed at different stages by S19. Three characteristic stages are considered: a. when the MS is {\it{grazing}} the RG, b. when it is starting to {\it{plunge-in}} through the envelope of the RG, and c. when the MS is well within the CE. We consider constantly powered and self-regulated jets, i.e., with a kinetic luminosity depending on the amount of matter accreting on the secondary, as described below.

The paper is organised as follows. In Section~\ref{sec:setup} we describe the numerical setups and the details of each model. In Section~\ref{sec:results}, we discuss the global evolution, mass accretion and specific angular momentum rates during the grazing and the plunge-in stages. In Section~\ref{sec:discussion}, we discuss our main results and analyse the effects that successful or choked jets in the CE produce. In Section~\ref{sec:conc}, we present our conclusions.

\section{Grazing and plunge-in setups and models}
\label{sec:setup}
\begin{figure}
   \includegraphics[width=0.49\textwidth]{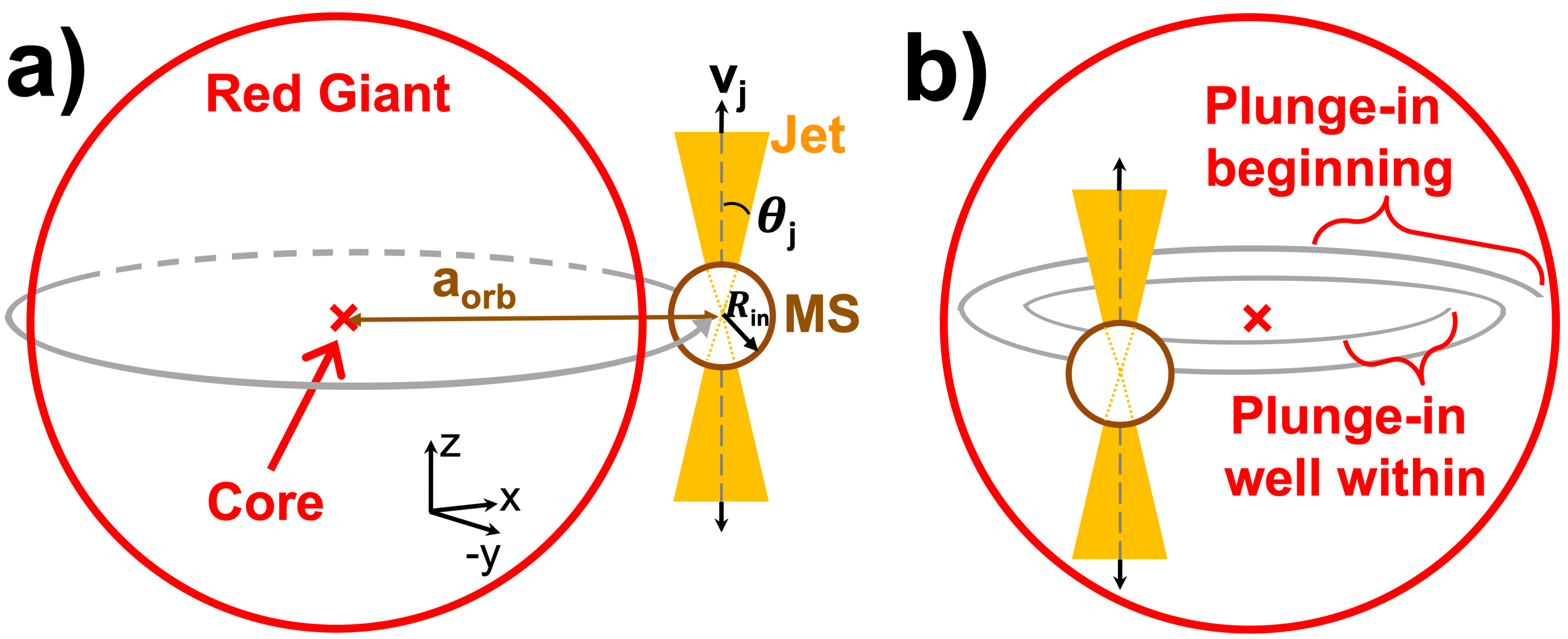}
   \caption{Scheme of the characteristic moments of the evolution of a MS-RG system followed in this study. Panel a) shows the {\it{grazing}} stage; and panel b) the {\it{plunge-in}} stage, i.e. when the MS moves through the CE of the RG (we study when the MS has just begun the {\it{plunge-in}} as well as when the MS is well within the CE). We study self-regulated as well as constantly-powered jets, as well as poorly- and highly-powered jets. They are launched perpendicular to the equatorial plane from the inner boundary ($R_{\rm{in}}$, set at the orbital separation $a_{\rm orb, init}$), and have an opening angle $\theta_j$ and velocity $v_j$.}
   \label{fig1}
\end{figure}

Figure~\ref{fig1} shows a schematic of the characteristic moments of the evolution of a MS-RG system followed in this study. The MS star launches two perpendicular jets as it spirals around the core of the RG. The phases which we study are the {\it{grazing}} stage (panel {\it a}) and the {\it{plunge-in}} stage (panel {\it b}). For the latter, we study the specific moments when the {\it{plunge-in}} has commenced, and when the MS is well within the CE.The jets may be self-regulated by the accretion rate that crosses the inner boundary ($R_{\rm in}$), or may be constantly powered. 

The initial configuration of the MS-RG system during the grazing, beginning of the plunge-in (pib), and when the MS star is well within the plunge-in (piww) stages were taken from S19. Also, the orbital separation between the MS and the core of the RG ($a_{\rm orb}$) and hydro boundary conditions are set to the S19 values at every time step. Specifically, we base our study on simulation \#5, in which the MS star orbits around the RG and does {\bf{not}} launch jets as it enters the CE \footnote{Our simulations start with the companion at the surface of the primary, however previous studies \citep{iaconi2017, shiber2019} showed that for a larger initial separation, the envelope is more inflated at the point when the companion starts spiraling-in into the envelope, and therefore we expect it would be easier for the jets to be successful for larger initial separation. We chose simulation \#5 of \citet{shiber2019} as a base to our simulations primarily because the orbital evolution in simulations \#5 (without jets) and \#6 (with jets) is almost identical in the first ~50 days and thus it is easier to compare our results with theirs.}. The MS star has mass $m_{\rm MS} = 0.30~\msun$, and radius $R_{\rm MS} \sim 0.30~\rsun$\footnote{The MS star is smaller than the inner boundary ($R_{\rm MS} < R_{\rm{in}}$).}. The RG has a total mass of $m_{\rm RG} = 0.88~\msun$, core mass $m_{\rm RG,c} = 0.39~\msun$, and initial radius of $R_{\rm RG} = 83~\rsun = 5.77 \times 10^{12}$~cm. The initial orbital separation of the grazing-envelope stage is $a_{\rm orb, init} \simeq R_{\rm RG}$ (which corresponds to $t \equiv t_0=0$~days), at the pib stage is $0.76 R_{\rm RG}$ ($t=35$~days~$\equiv t_{pib}$), and at the piww is $0.48 R_{\rm RG}$ ($t=52$~days~$\equiv t_{piww}$). The evolution of the orbital separation with time is shown in Figure 6 of S19. The change in the orbital separation in each simulation is relatively small. We neglect the (small) change in the orbital velocity caused by the movement of the frame of reference at each time step.

Our simulations are performed in the rest frame of the MS star (non-inertial frame), thus, a wind (due to the motion of the MS star around the RG), and the Coriolis and Centrifugal forces are considered. The gravitational force is estimated at each time step directly from the numerical simulation. The gravitational effects of the RG star are taken as radial and consider the mass enclosed within the orbital separation in the momentum and energy conservation equations. The mass of the MS is set as as point mass in its correspondent position (the amount of mass which it accretes during our integration time is negligible compared to that of the MS star, thus the mass of the MS star is maintained constant). We do not include self-gravity, but the interaction between the gas and the companion star is followed precisely in the simulation by S19, which includes self-gravity, and which our simulations are based upon.

In each of the MS-RG stages, we follow the evolution of self-regulated or constantly powered jets, with different powering,  during $10^6$~s (11.6 days). We assume that the material crossing the inner boundary is either accreted onto the stellar surface or ejected as jets. The jets are injected perpendicularly from an inner boundary located at $R_{\rm{in}} = 10^{11}$~cm. Each jet has an opening angle $\theta_j = 30^\circ$, a velocity $v_j  = 438$~km~s$^{-1}$ and is defined inside a conical region with $R<R_{\rm{in}}$ and $\theta < \theta_j$. The material inside the inner boundary is fully removed at each time step (with density pressure and velocity rewritten to their initial value within the inner boundary\footnote{As the gravitational force diverges in a region which is outside the computational box, we do not include any  gravitational softening.}).

The density of each jet is defined as $\rho= \dot{M}/(4\pi(1-\cos{\theta_j}) R_{\rm{in}}^2 v_j$. The jet pressure is defined as $P=n k_B T$, where $n=\rho/\mu m_H$ is the number density, $\mu$ is the mean molecular weight, $m_H$ the proton mass, and we take a temperature of $T/\mu=10^6$ K. The jets are injected without angular momentum (although some angular momentum, as described in the paper, is present at the inner boundary).

The jets were powered by a fraction of the mass accretion rate ($\eta_{\rm sr}$ for the self-regulated jets and $\eta_{\rm c}$ for the constantly powered jets). The luminosity of the self-regulating jets was $L_j = \eta_{\rm sr} \dot{M}_{\rm acc} v_j^2$, where $\dot{M}_{\rm acc}$ is the mass accretion rate that crosses $R_{\rm{in}}$. Meanwhile, the luminosity of the constantly powered models was $L_j = \eta_{\rm c} \dot{M}_{\rm BHL} v_j^2$, where $\dot{M}_{\rm BHL}$ is the assumed Bondi-Hoyle-Littleton rate \citep{hoylelyttleton1939, bondihoyle1944} at the RG surface in the simulations of S19\footnote{The $\dot{M}_{\rm BHL}$ is estimated at the surface of the accretor, the MS ($R\sim 0.3 \rsun$); the accreting wind velocity and density are estimated from the Keplerian velocity (we assume there was no prior tidal synchronization) and the density profile of the surface of the donor star, when the MS star is near the surface of the RG star $\dot{M}_{\rm BHL} \approx 10^{25}$g~s$^{-1}$.}. We explore accretion efficiency values up to 0.70 (the high-$\eta$ values are used to explore the most extreme cases).

The simulations were carried out by employing the 3D, adaptive mesh refinement, HD code {\it Mezcal} \citep{decolle2012}. The numerical domain in all stages covered a large fraction of the CE ($\Delta X = \Delta Y = 0.76 R_{\rm RG} = 4.42 \times 10^{12}$~cm~$, \Delta Z = 2 \Delta X$). Five levels of refinement, corresponding to a maximum resolution of $\delta X = \delta Y = \delta Z = 4.90 \times 10^{-2} R_{\rm RG} = 3.44 \times 10^{9}$~cm are used. A fixed spherical region centered in the MS star and with a radius of $0.15\Delta Z$ is allowed to refine to the maximum resolution. The maximum levels of refinement drop for regions outside $0.3\Delta Z, 0.45\Delta Z, 0.6\Delta Z$. Each jet was injected in $\approx$ 700 cells at the solid angle produced by $R=R_{\rm{in}}$ and $\theta \leq \theta_j$ in its respective hemisphere (the reminder of the $R=R_{\rm{in}}$ surface, $\approx 10^4$cells, has an outlfow condition). The inner part of the RG and the motion of the MS star were taken into account by setting the $X_{\rm min}$ and the $Y_{\rm max}$ with the values coming from the large scale simulation of S19. All other boundaries were set with outflow boundary conditions.

The mass accretion rate ($\dot{M}$) and angular momentum rate ($\dot{J}$), which cross $R=R_{\rm{in}}$, are saved before the material is removed from the computational domain). The specific angular ($J_{\rm sp}$) momentum is obtained by $J_{\rm sp} =\dot{J}$/$\dot{M}$. The integration time for each simulation is t$_{\rm int}= 10^{6}$~s, and the Courant number is set at $n_{\rm C} = 0.1$. We verify that, in the absence of jets, the common envelope remains in quasi-hydrostatic equilibrium for the duration of the simulation. In the self-regulating jet models we have assumed that the material that crosses $R_{\rm{in}}$ immediately alters the powering of the jets. The latter is not true since the material requires a time-lag ($t_{\rm{lag}}$) in order to reach the surface of the MS star. This however does not modify the main results since including a $t_{\rm{lag}} >0$ does not affect the global outcome \citep[see for example, ][]{lc2019}.

Thirty models are followed, see Table~\ref{table1}. The models are labeled according to the MS-RG stage (``g'' for {\it{grazing}}, ``pib'' for the beginning of {\it{plunge-in}} phase, and ``piww'' for when the MS star is well within the CE); according to whether the jets were self-regulated (sr), constant (c), or if no jets were present (nj); and according to the $\eta$ value. The models which have successful jets are indicated by a check-mark.

\begin{table}
\begin{center}
\begin{tabular}{ccccc}
  \hline
  Model &    Stage      & Jets  &  $\eta_{\rm sr}$ or $\eta_{\rm c}$ &  Outcome  \\
  \hline
  g-nj-0.00  &  {\it{grazing}}   &     NJ     & 0.00 & $X$  \\
  g-sr-0.02  &  {\it{grazing}}   &     SR     & 0.02 & $X$  \\
  g-sr-0.05  &  {\it{grazing}}   &     SR     & 0.05 & $X$  \\
  g-sr-0.10  &  {\it{grazing}}   &     SR     & 0.10 & $\checkmark$  \\
  g-sr-0.30  &  {\it{grazing}}   &     SR     & 0.30 & $\checkmark$  \\
  g-sr-0.50  &  {\it{grazing}}   &     SR     & 0.50 & $\checkmark$  \\

  g-c-0.02  &  {\it{grazing}}   &     C     & 0.02 & $\checkmark$  \\
  g-c-0.05  &  {\it{grazing}}   &     C     & 0.05 & $\checkmark$  \\

  pib-nj-0.00  &  {\it{PI}}-beginning   &     NJ     & 0.00 & $X$  \\
  pib-sr-0.02  &  {\it{PI}}-beginning   &     SR     & 0.02 & $X$  \\
  pib-sr-0.05  &  {\it{PI}}-beginning   &     SR     & 0.05 & $X$  \\
  pib-sr-0.10  &  {\it{PI}}-beginning   &     SR     & 0.10 & $X$  \\
  pib-sr-0.30  &  {\it{PI}}-beginning   &     SR     & 0.30 & $X$  \\
  pib-sr-0.50  &  {\it{PI}}-beginning   &     SR     & 0.50 & $\checkmark$  \\

  pib-c-0.02  &  {\it{PI}}-beginning   &     C     & 0.02 & $X$  \\
  pib-c-0.05  &  {\it{PI}}-beginning   &     C     & 0.05 & $X$  \\
  pib-c-0.10  &  {\it{PI}}-beginning   &     C     & 0.10 & $X$  \\
  pib-c-0.30  &  {\it{PI}}-beginning   &     C     & 0.30 & $\checkmark$  \\
  pib-c-0.50  &  {\it{PI}}-beginning   &     C     & 0.50 & $\checkmark$  \\

  piww-nj-0.00  &  {\it{PI}}-well within   &     NJ     & 0.00 & $X$  \\
  piww-sr-0.05  &  {\it{PI}}-well within   &     SR     & 0.05 & $X$  \\
  piww-sr-0.10  &  {\it{PI}}-well within   &     SR     & 0.10 & $X$  \\
  piww-sr-0.30  &  {\it{PI}}-well within   &     SR     & 0.30 & $X$  \\
  piww-sr-0.50  &  {\it{PI}}-well within   &     SR     & 0.50 & $X$  \\
  piww-sr-0.70  &  {\it{PI}}-well within   &     SR     & 0.70 & $X$  \\
  
  piww-c-0.05  &  {\it{PI}}-well within   &     C     & 0.05 & $X$  \\
  piww-c-0.10  &  {\it{PI}}-well within   &     C     & 0.10 & $X$  \\
  piww-c-0.30  &  {\it{PI}}-well within   &     C     & 0.03 & $X$  \\
  piww-c-0.50  &  {\it{PI}}-well within   &     C     & 0.50 & $X$  \\
  piww-c-0.70  &  {\it{PI}}-well within   &     C     & 0.70 & $X$  \\
  \hline
\end{tabular}
\end{center}
\caption{Stage and jet characteristics of each model.  The models are labels according to: a) the MS-RG stage (g for {\it{grazing}}, pib for when the {\it{plunge-in}} is beginning, and piww for when the MS star is well within the CE), b) if the jets were self-regulated (sr), or constant (c), or if no jets were present (nj), and c) the efficiency value. The used acronyms are {\it{PI}} for {\it{plunge-in}}, SR for self-regulated, C for constant. The models which have successful jets are indicated by a check-mark, choked jets (or no jets) are indicated by $X$.}
\label{table1}
\end{table}

\section{Results}
\label{sec:results}
In this section we show the stellar envelope morphology and its evolution as a result of a MS star which launches two jets perpendicular to the orbital plane either when it grazes or when it plunges-in through the envelope of a RG. We also discuss the differences between having a pair of self-regulated jets vs constantly powered jets.

\subsection{Grazing stage}
\label{sec:grazing}
We first describe the expected outcomes from when the envelope of the RG star is grazed by a MS star. The MS star can launch jets with either enough ram pressure to overcome the stellar envelope material which surrounds the MS star and to break out of the CE (which we term as ``successful"), or jets which do not overcome the ram pressure of the envelope material and are not able to break out of the CE (``choked"). In column 5 of Table~\ref{table1} we list whether the jets from each model are successful or choked at the end of each simulation.

\begin{figure*}
   \includegraphics[width=0.95\textwidth]{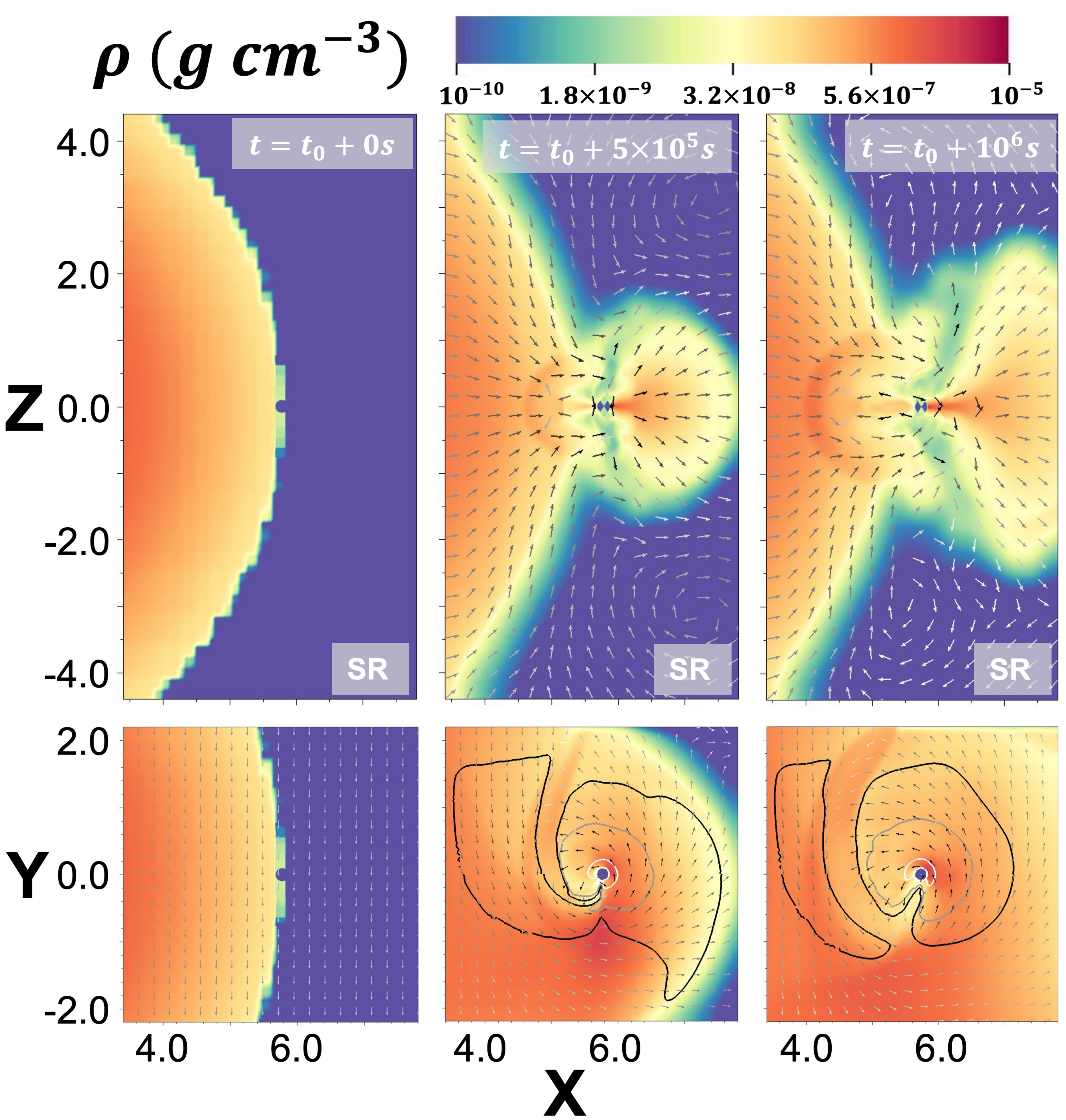}
   \caption{Density map plots showing the evolution of a successful self-regulated (SR) pair of jets with $\eta$ = 0.10 (model g-sr-0.10), when the MS star is {\it{grazing}} the envelope of the RG. The meridional (upper) and equatorial (lower) planes are shown at characteristic times ($t=t_0+0s$, $t_0+5\times10^5s$, $t_0+10^6s$, with $t_0=0$~days). The initial orbital separation at $t_0$ is $a_{\rm orb, init} =  R_{\rm RG}$. The axis are in units of $10^{12}$~cm. The velocity field is indicated by the arrows (white: v=$10^5$~cm~s$^{-1}$, grey: v=$5\times10^5$~cm~s$^{-1}$, black: v=$10^7$~cm~s$^{-1}$). The magnitude of the component of the velocity in the equatorial plane is indicated by the isocontours (black: $v_{\rm eq}=6.0 \times 10^{6}$~cm~s$^{-1}$, grey: $v_{\rm eq}=8.6 \times 10^{6}$~cm~s$^{-1}$, and white: $v_{\rm eq}=1.7 \times 10^{7}$~cm~s$^{-1}$). The left panels show the resolution at $t=t_0$ taken from S19 (with $\delta X \sim$7$\times10^{10}$~cm). As soon as $t>t_0$, the AMR adapts to finer resolutions ($\delta X \sim$3.44$\times10^{9}$~cm). The envelope was stable without the presence of the MS and increased its size by <5\% of its initial value during the integration time.}
   \label{fig2}
\end{figure*}

We find that in the grazing stage successful jets are easily launched (they require $L_j \sim 10^{37}$~erg~s$^{-1}$), and no clear disk forms around the MS star. Figure~\ref{fig2} shows the results for model with $\eta=0.10$ (g-sr-0.10). The meridional (upper panels) and equatorial (lower panels) planes with the density-map plots and the velocity field at three different times are shown. 
The left panels show the initial condition ($t=t_0+0$~s, with $t_0=0$~days) when the RG is grazed by the MS star and the jets are about to be launched. The envelope of the RG has density values which range from $\sim 10^{-8}$~g~cm$^{-3}$ in the outskirts of the envelope to $\sim 10^{-5}$~g~cm$^{-3}$ near the core. The ambient medium around the RG is static, however since the reference system is placed in the MS, thus, the {\it{grazing}} motion of the MS around the RG star produces the initial $v_y \neq 0$ clearly seen in the equatorial plane. 
The middle panels show a snapshot ($t=t_0+5\times10^5$~s) when the jets are successful. Each jet propagates through a low-density channel formed within the bulge (centered in the MS star). The bulge has densities of order $\sim 10^{-8}-10^{-6}$~g~cm$^{-3}$. The jets have density and velocity values of order $\sim 10^{-9}$~g~cm$^{-3}$ and $10^7$~cm~s$^{-1}$ respectively. A wind (with $\sim 10^{-8}$~g~cm$^{-3}$ and $v\sim10^6$~cm~s$^{-1}$) is produced. Meanwhile, a fraction of the material that is being accreted by the gravitational pull of the MS star circulates in the equatorial plane around the MS star (with $\sim10^7$~cm~s$^{-1}$).

The panels on the right show the configuration at the end of the {\it{grazing}} stage simulation ($t=t_0+10^6$~s). The jets are successful and the bulge is still present. The density of the jets and their velocity are practically unchanged as the jets are now further away from the MS star. The bulge basically maintains the same properties (density, equatorial morphology flow, and wind) and has now expanded. By the end of the integration time the initial orbital separation ($a_{\rm orb}=R_{\rm RG}=83~\rsun$) diminishes to $a_{\rm orb}=0.99~R_{\rm RG}=82.4~\rsun$. The self-regulated jets with $\eta>0.10$ had enough ram pressure for the jets to also be successful and follow the same global evolution. 

\begin{figure*}
   \includegraphics[width=0.95\textwidth]{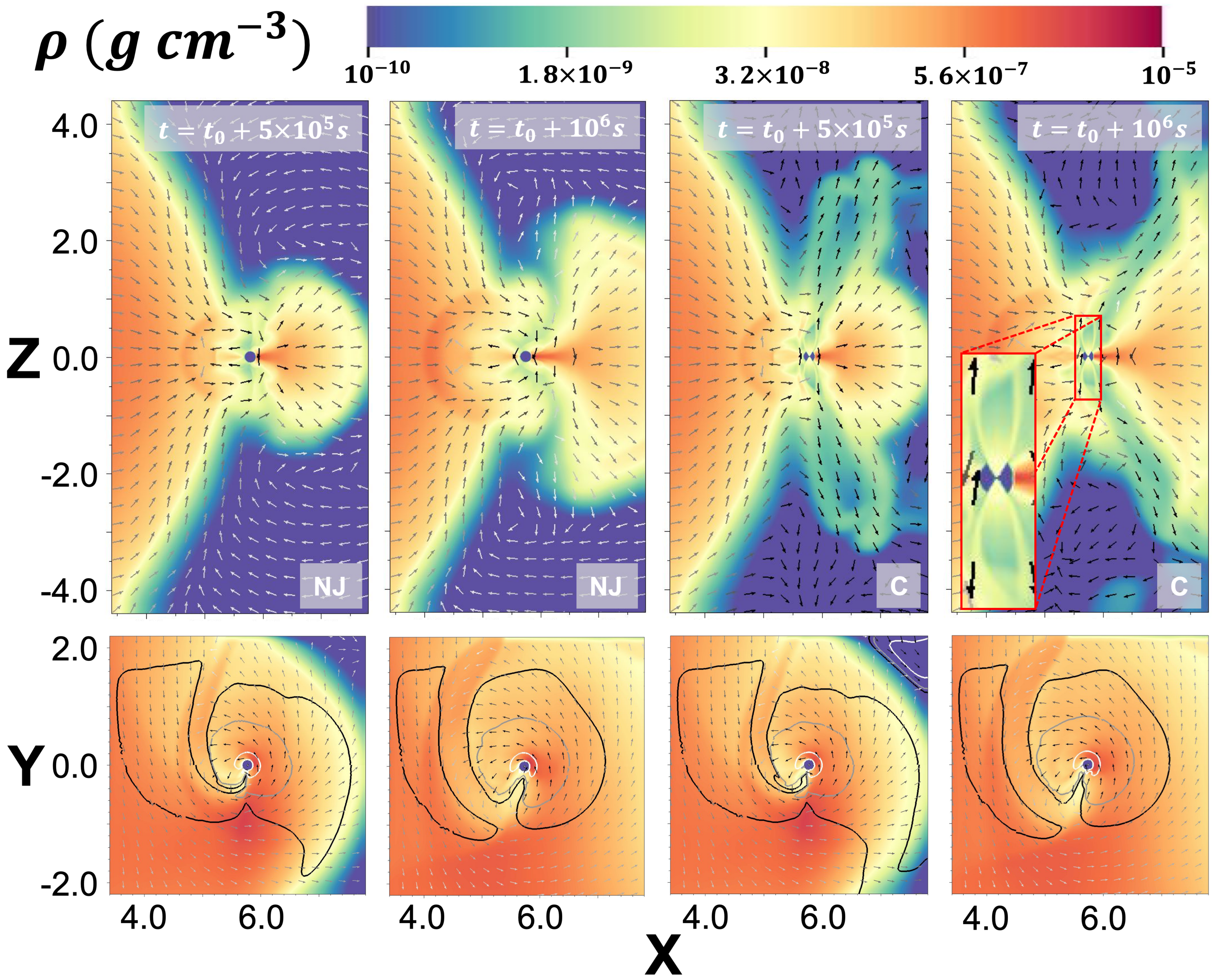}
   \caption{Density map plots showing the evolution during the grazing stage when no jets (NJ) are launched (model g-nj-0.00, two leftmost panels) and when the pair of successful jets are constantly powered (model g-c-0.02, two rightmost panels). In each case the configuration at the characteristic times $t_0+5\times10^5$~s and $t_0+10^6$~s (with $t_0=0$~days) are shown. The initial setup ($t=t_0+0s$) is not shown as it is the same as for the self-regulated case (left panel in Figure~\ref{fig2}). The axis units, $a_{\rm orb, init}$, velocity field magnitude, and isocontour values are the same as in Figure~\ref{fig2}. The inset in the upper-right panel shows a zoom of the inner region where the jets are being launched.}
   \label{fig3}
\end{figure*}

In Figure~\ref{fig3} we show the evolution of the MS-RG system during the {\it{grazing}} stage when no jets (NJ) are launched (model g-nj-0.00, two left panels) and when the pair of successful jets are constantly (C) powered (with $\eta$ = 0.02, model g-c-0.02, two right panels). The meridional and equatorial planes with the density-map plots and the velocity field at two different times are shown ($t_0+5\times10^5$~s and $t_0+10^6s$, the initial setup is not shown as it is the same as for the self-regulated case). The two left panels show how in the NJ model the bulge forms without the presence of a jet and is basically the same as that of the SR model shown in Figure~\ref{fig2} (as well as the accreted material which circulates around the MS star in the equatorial plane). A low-density channel is still formed within the bulge, but now with material falling towards the MS star. The two right panels show the evolution of a pair of successful and constantly powered jets (model g-c-0.02). The bulge morphology, the jet density and velocities are similar to those of the SR models. The jets initially move nearly vertically (see the zoomed inset in the upper-right panel). As the jets interact with the bulge, they are deflected and flow through the low-density channel (see the velocity field). Constantly-powered jets with $\eta>0.02$ present the same global morphology, evolution, and are also successful. The equatorial flow morphology is not affected by the presence or nature of the jets (SR or C) given that the jets are perpendicular to the XY axis.

\begin{figure}
   \includegraphics[width=0.47\textwidth]{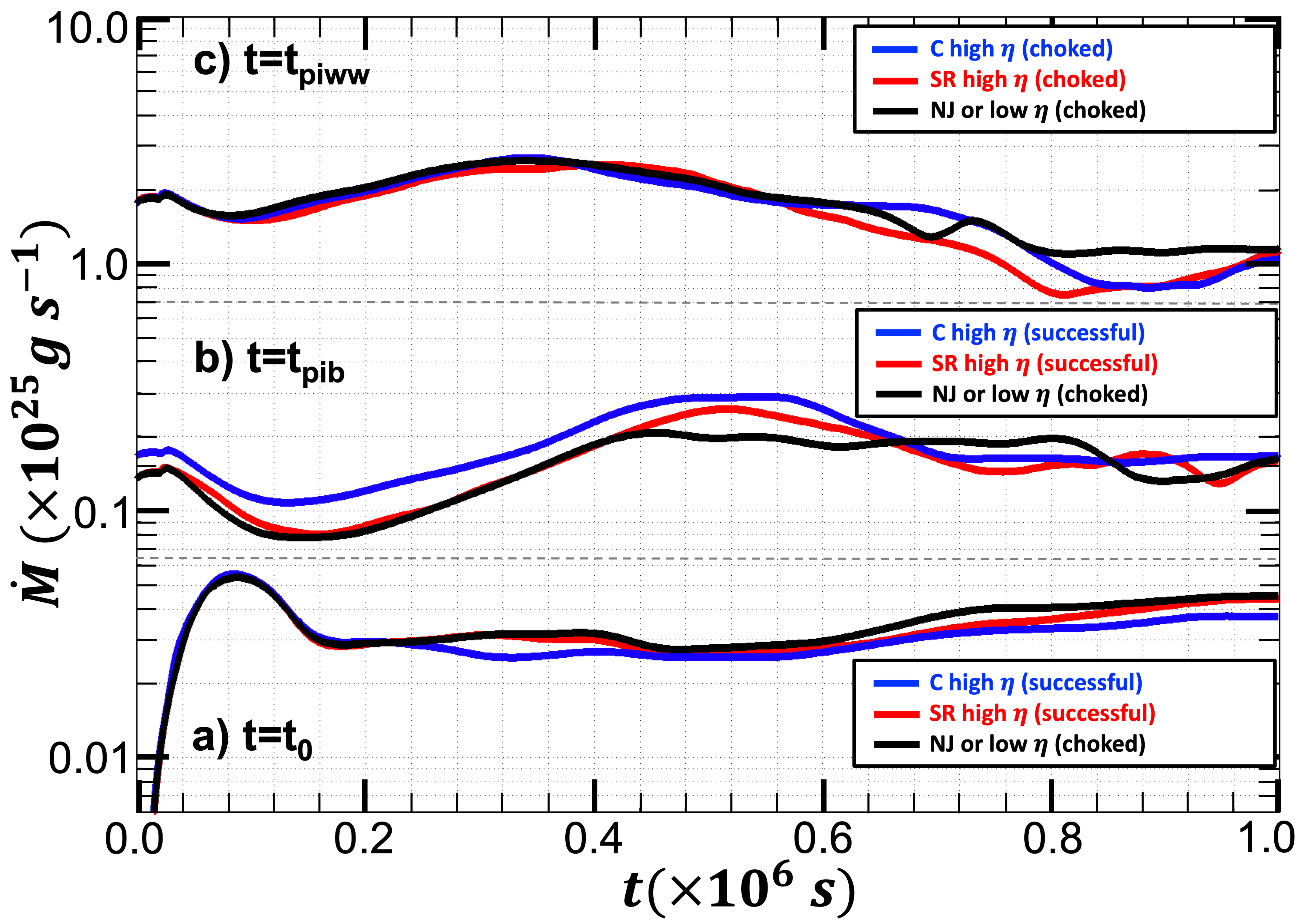}
      \caption{Mass accretion rates at $R_{\rm in}$ during: the {\it{grazing}} stage, $a) \ t=t_0$; the beginning of the {\it{plunge-in}} stage, $b) \ t=t_{pib}$; and when the MS star is well within the CE, $c) \ t=t_{piww}$. Self-regulated jet models with high $\eta$ are indicated in blue, constantly powered jets with high $\eta$ in red, and low $\eta$ choked jets in black. The horizontal dashed lines only separate cases a), b), and c) and have no physical meaning.}
   \label{fig4}
\end{figure}

\begin{figure}
   \includegraphics[width=0.47\textwidth]{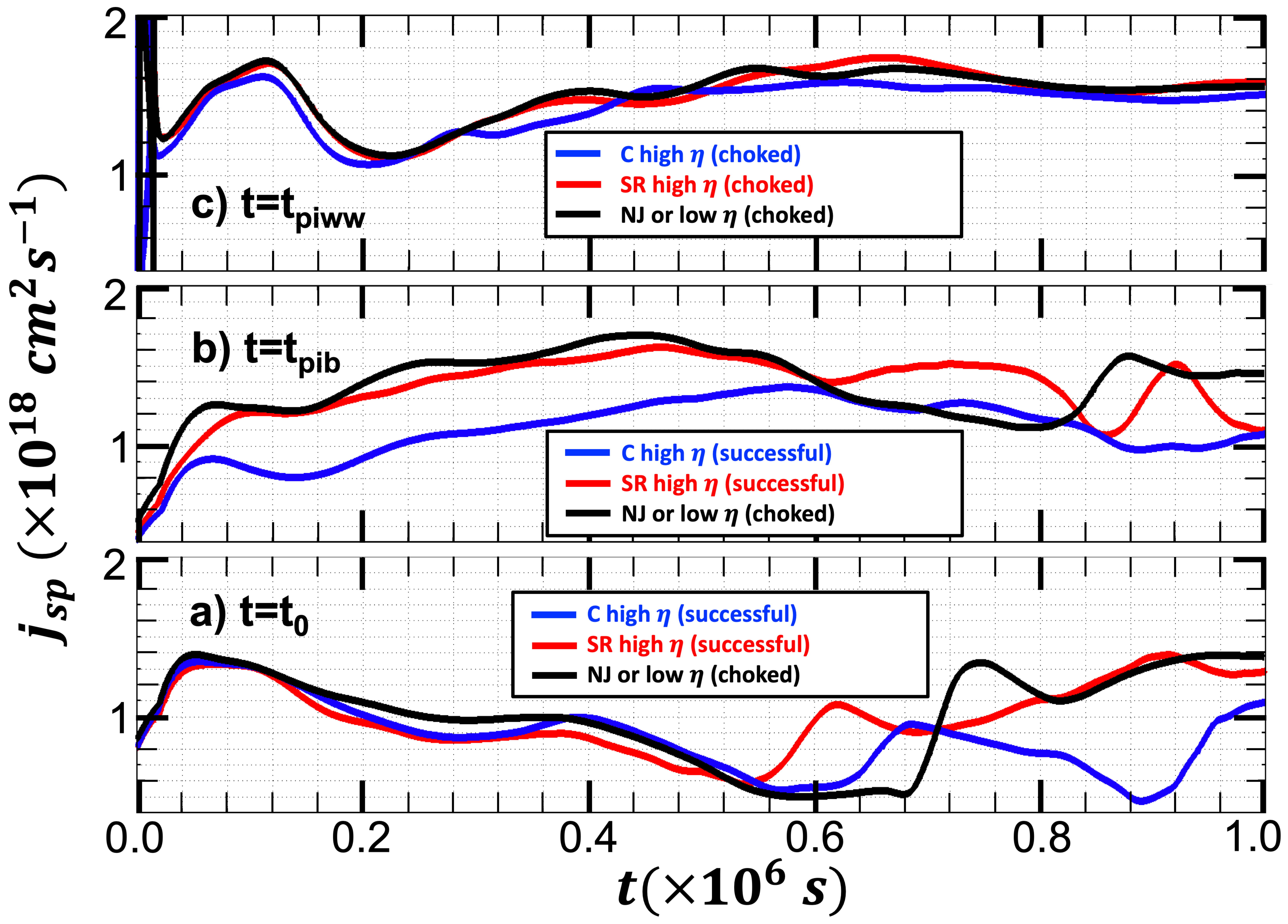}
      \caption{Specific angular momentum at $R_{\rm in}$ for the models indicated in Figure~\ref{fig4}.}
   \label{fig5}
\end{figure}

The mass-accretion rate and specific angular momentum for the successful jet models and the choked (or NJ) models, during the grazing stage ($t=t_0$), are shown in Figure~\ref{fig4} and Figure~\ref{fig5}, respectively. Specifically, we show the mass accretion and specific angular momentum at $R_{\rm{in}}$ ($\dot{M}_{\rm{in}}$, $J_{\rm sp, in}$, respectively) for the successful SR model g-sr-0.10 (termed as SR high $\eta$), the successful C model g-c-0.02 (C high $\eta$), and the NJ (model g-nj-0.00), i.e. SR and C models with low $\eta$ which choke (NJ or SR-C low $\eta$). The $\dot{M}_{\rm{in}}$ of the successful self-regulated jets tends to be marginally lower than that of the successful constantly-powered jets. The choked-jet models tend to have slightly higher accretion rates. The mass accretion rate is nearly identical in all the cases (this is, either when no jets were launched, when they were choked, or when they were successful). Independently of the nature of the jets (choked or not), the mass accretion at the inner boundary was within $\dot{M}_{\rm{in}} \approx 2-6 \times 10^{23}$~g~s$^{-1}$. Thus, the self-regulated jets during the {\it{grazing}} stage had a $L_j \sim 10^{36}-10^{37}$~erg~s$^{-1}$ (depending on the $\eta$ value), which is about an order of magnitude below the constantly-powered-jet models luminosity. As for the accretion rate, in the choked jet models the $J_{\rm sp, in}$ tends to be moderately higher than that of the successful jets. Still, there is no clear difference in the $J_{\rm sp, in}$ between the successful self-regulated jets, the constantly powered successful jets, and the choked jet models. For all cases $J_{\rm sp, in} \approx 0.4-1.4 \times 10^{18}$~cm$^{2}$~s$^{-1}$.

\subsection{Plunge-in stage}
\label{sec:plunge-in}
We next discuss the global evolution of the system as the MS star {\it{plunges-in}} through the envelope of the RG. The jets are launched once the MS is within the CE, i.e., either during the pib or piww stages. Table~\ref{table1} lists whether the jet models at each stage are successful or choked at the end of the integration time. 

In the large scale global simulation of S19 which is used as our initial configuration, the plunging-in of the MS star heats the envelope at the MS star surroundings. As a result, the envelope is inflated (and remains mostly bound), and an equatorial outflow is produced. In addition, envelope mass accumulates behind the path of the MS and concentrates around the equatorial plane in a bulge-like structure. Later, when the MS star further plunges-in, both the MS star and the bulge are engulfed inside the envelope (see left-most panels of figures 6 and 7). Because initially the envelope in the companion's vicinity concentrates around the equatorial plane the jets will have less layers of envelope to interact with in the vertical direction, which will ease the launching of successful jets, while in the piww stage when the bulge is immersed inside the envelope there is a greater chance that the jets would be choked.

\begin{figure*}
   \includegraphics[width=0.95\textwidth]{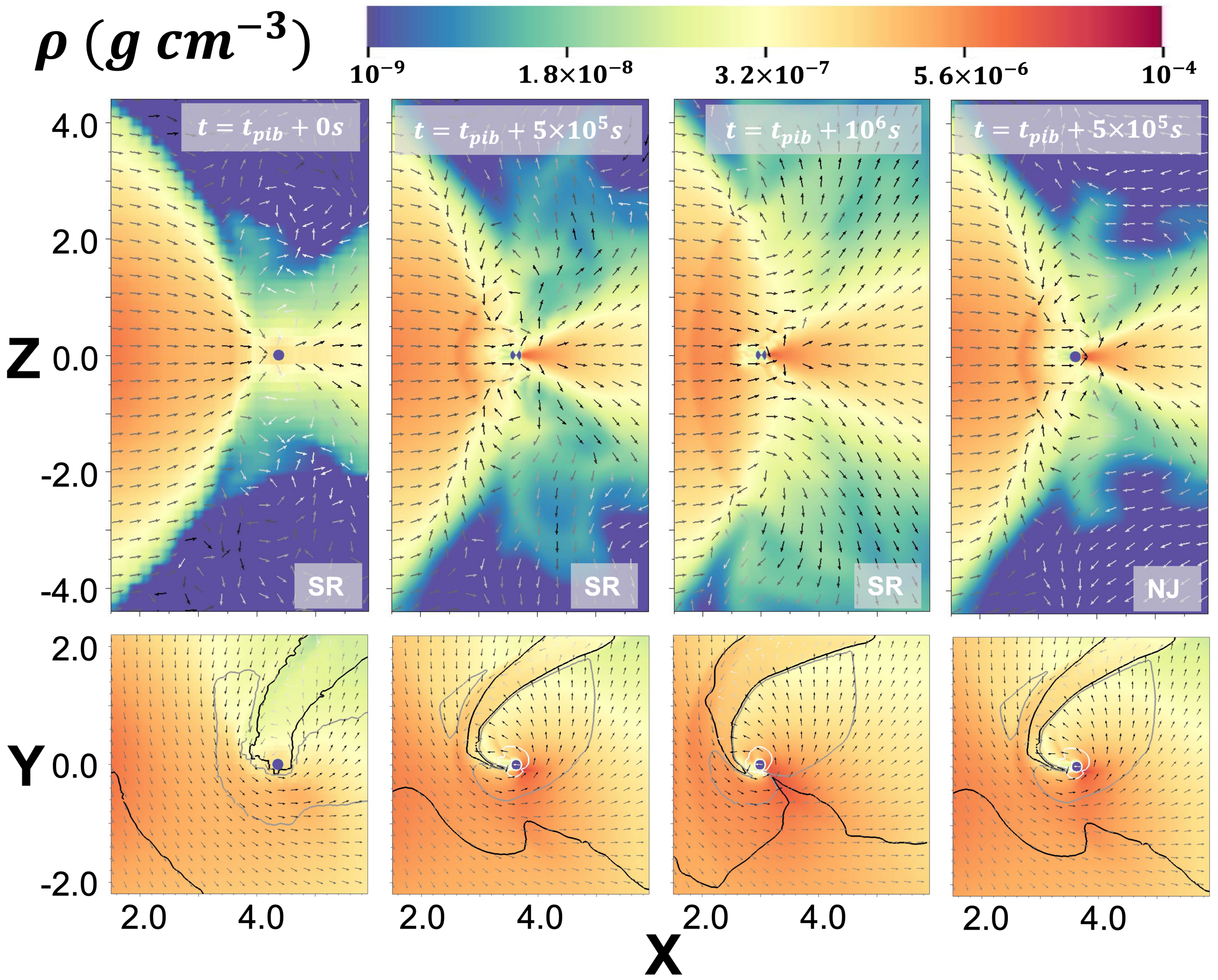}
   \caption{Density map plots showing the evolution of a successful SR jets with $\eta$ = 0.50 (model pib-sr-0.50) for when the MS star is starting the {\it{plunge-in}} phase through the envelope of the RG (where $t_{pib}=35$~days, and $a_{\rm orb, init} = 0.76 R_{\rm RG}$). The meridional (upper) and equatorial (lower) planes are shown at characteristic times ($t=t_{pib}+0s$, $t_{pib}+5\times10^5s$, $t_{pib}+10^6s$). A snapshot of model pib-nj-0.00 is also included (upper and lower right panels). The axis units, velocity field magnitude, and velocity isocontours are the same as Figure~\ref{fig2}.}
   \label{fig6}
\end{figure*}

In the pib stage successful jets require higher $\eta$ powering (compared to the grazing stage). Figure~\ref{fig6} shows the meridional and equatorial density maps and velocity field evolution (upper and lower panels, respectively) for the SR model with $\eta=0.50$ during the pib stage (model pib-sr-0.50). The orbital separation in this stage changes from $a_{\rm orb}=0.76~R_{\rm RG}$ to $a_{\rm orb}=0.49~R_{\rm RG}$. The three leftmost panels show the initial condition and two subsequent times ($5\times10^5$~s and $10^6$~s) exhibiting how the jets are able to drill through the RG. For comparison, a snapshot of the model where no jets are launched is also included (pib-nj-0.00 at $5\times10^5$~s, rightmost panel). The morphology of the bulge and the region accreting onto the secondary star are very similar in the cases with/without jets. The bulge, composed once more by a wind and equatorial material (with densities $\sim 10^{-6}$~g~cm$^{-3}$), is present, and no clear disk forms around the MS star. The jets have density and velocity values of order $\sim 10^{-8}$~g~cm$^{-3}$ and $10^7$~cm~s$^{-1}$ respectively. Since the material that surrounds the MS star is denser than in the {\it{grazing}} stage, the jets must be more powerful in order to be successful. For the self-regulated jets to be successful the $\eta$ efficiency must be much larger ($\eta\ge0.50$) than that in the {\it{grazing}} stage. Thus, the jets now require a luminosity $\sim 10^{39}$~erg~s$^{-1}$ in order to be successful. Constantly powered jet models with $\eta>0.30$ value are also successful, and present a similar global morphology and evolution.

\begin{figure*}
   \includegraphics[width=0.95\textwidth]{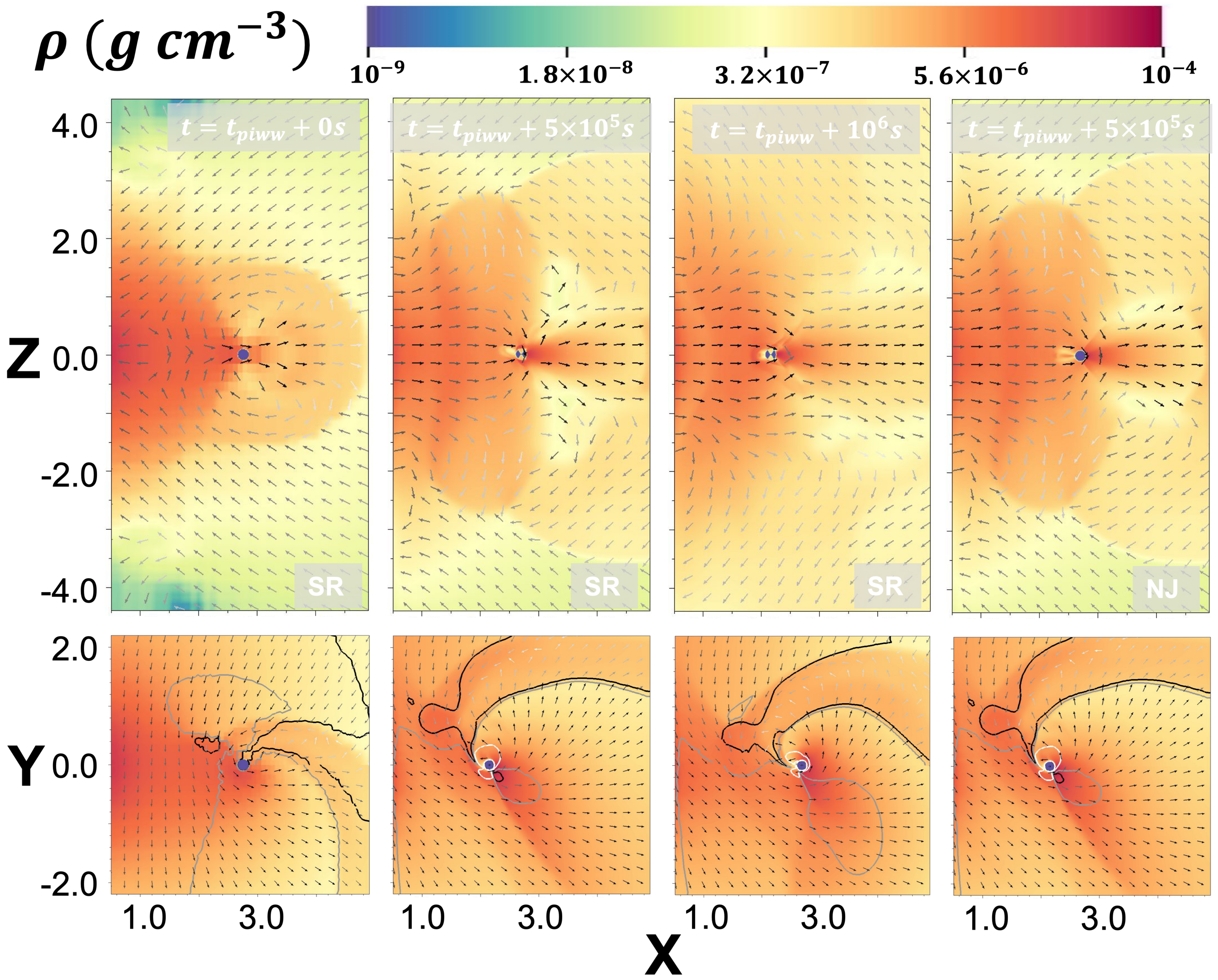}
      \caption{Same as Figure~\ref{fig7} for when the MS star is well within the RG ($t_{piww}=52$~days, nd $a_{\rm orb, init} = 0.48 R_{\rm RG}$).}
   \label{fig7}
\end{figure*}

In the piww stage all jets, independently of their nature or $\eta$ efficiency, are choked. Figure~\ref{fig7} shows the meridional and equatorial density maps and velocity field evolution (upper and lower panels, respectively) for the SR model with $\eta=0.70$ during the piww stage (model piww-sr-0.70). The orbital separation in the piww stage changes from $a_{\rm orb}=0.48~R_{\rm RG}$ to $a_{\rm orb}=0.47~R_{\rm RG}$. The three leftmost panels show the initial condition and two subsequent times when the MS is well within the RG. Also, a snapshot of the model where no jets are launched is also included (piww-nj-0.00, rightmost panel). The high density which surrounds the MS star is such that the ram pressure of the jet is not able to drill through the core of the RG, and thus the jets are always choked (independently if they are SR or C powered). The jets manage to initially expand outside the inner boundary but are quickly choked ($\sim 10^5s$) inside the bulge (which is slightly modified by the presence of the choked jet). The equatorial morphology flow is not affected by the presence of jets and no disk forms around the MS star.

The $\dot{M}_{\rm{in}}$ and $J_{\rm sp, in}$ for the successful SR or C jet models and the NJ or choked models, during the pib and piww stage ($t=t_{\rm{pib}}$, and $t=t_{\rm{piww}}$), are shown in Figure~\ref{fig4} and Figure~\ref{fig5}, respectively. In the pib stage the we show model pib-sr-0.50 as the successful SR model, model pib-c-0.30 as the successful C model, and pib-nj-0.00 for the choked model. In the piww stage the we show models piww-sr-0.70, piww-c-0.70, and piww-nj-0.00. During the pib the accretion rates are within $\dot{M}_{\rm{in}} \approx 0.7-4.0 \times 10^{24}$~g~s$^{-1}$, which is about an order of magnitude larger than in the {\it{grazing}} stage. Meanwhile, at the piww stage, where all the jets were choked, the accretion rate was $\sim$ two orders of magnitude larger than in the {\it{grazing}} stage, this is, $\dot{M}_{\rm{in}} \approx 0.7-3.0 \times 10^{25}$~g~s$^{-1}$. The $J_{\rm sp, in}$ during the pib stage is $J_{\rm sp, in} \approx 0.4-1.7 \sim 10^{18}$~cm$^{2}$~s$^{-1}$, and at the piww stage is $J_{\rm sp, in} \approx 1.0-1.8 \sim 10^{18}$~cm$^{2}$~s$^{-1}$.

\section{Discussion}
\label{sec:discussion}
We have studied the dynamics and propagation of jets launched from a MS star moving through the envelope of a RG star. We followed, using 3D HD simulations, a set of jet models (either self-regulated or constantly powered, and with different kinetic luminosities) in three phases: when the MS is grazing the RG, when the MS star has just started to plunge-in through the CE, and when the MS star is well within the CE. Our numerical simulations show that jets can be choked or successful depending on the jet efficiency and on the evolutionary phase. {\it{Grazing}} jets and jets launched at the beginning of the plunge-in stage may be successful. Meanwhile, jets that are launched deep inside the CE are choked. As shown in Section \ref{sec:results}, it becomes increasingly difficult for the jets to break out through the accreting material as the MS star moves inwards into the CE. Once the jets are launched, independently of whether they are self-regulated or constantly powered, each jet will be successful if its ram pressure ($P_{\rm jet}$) is larger than the ram pressure of the accreting ambient material ($P_{\rm amb}$). The jet and the accreting material ram pressures are given by (see \citealt{mm2017} for further details)
\begin{eqnarray}
    P_{\rm jet} \simeq \rho_{\rm amb} v_\star^2 \frac{r_B^2}{r^2} \frac{2\eta v_j}{\theta_j^2 v_\star}\;, \qquad
    P_{\rm amb} \simeq \rho_{\rm amb} v_\star^2 \frac{r_B^2}{r^2}\;,
\label{eq1}
\end{eqnarray}
where $\rho_{\rm amb}$ is the density of the CE, $v_\star = \sqrt{G M(a)/a}$ is the velocity of the MS star within the RG (specifically at the orbital separation $a$, and taken as keplerian), $r_B = 2 G M_{\rm MS}/v_\star^2$ is the Bondi radius, $v_j$ the velocity of the jet, and $\theta_j$ the opening angle of the jet. From the ram pressure of the jets we can see that if they were faster, or if they were more collimated then smaller $\eta$ efficiencies would be required for the jets to be successful. More collimated jets have a much larger ram pressure (since $P_{\rm jet} \propto \theta_j^{-2}$), which may overcome the accreting material much more easily. In our case, if the opening angle of the jets was $\theta_j \approx 10^\circ$ (instead of the $\theta_j \approx 30^\circ$ used in our simulations) then the $\eta$ values for the jets to be successful would be $\sim$10 times smaller. 

Comparing the ram pressures, the condition for the jets to be successful $P_{\rm jet}>P_{\rm amb}$ implies that 
\begin{equation}
   a \gtrsim a_0 = 3\times 10^{12} {\rm cm}  \left(\frac{M(a)}{1 M_\odot}\right) \left( \frac{\theta_j}{30^\circ}\right)^4 \left( \frac{\eta}{0.1}\right)^{-2}  \left( \frac{v_j}{500 \; {\rm km} \; {\rm s}^{-1}}\right)^{-2}.
\label{eq2}
\end{equation} 
Equation~\ref{eq2} shows that outside a critical orbital separation $a_0$, jets are successful, while at smaller orbital separations the jets will be choked by the accreting material. The larger the efficiency, hence, smaller the orbital separation (and deeper within the CE) where the jets are successful. We also see that the value of $a_0$ strongly depends on the jet velocity, with larger velocity the jet ram pressure increases, thus making it easier for a jet to break out from deeper within the CE. The opening angle also noticeably modifies the critical orbital separation in which the jets would be successful (as $a_0 \propto \theta_j^4$). Hence, jets with small efficiency could produce successful jets if their opening angle was much smaller, at least in the intermediate ``plunge-in'' phase. In order for the jets to be successful, their ram pressure ought to be larger than that of the surrounding stellar envelope (both computed in the shock frame). This is true both in accreting systems and in static (not-accreting) flows, in which the ram pressure of the ambient medium corresponds to its inertia. The thermal pressure is not large enough to slow the jet down as the jet shock head is highly supersonic and in strong shocks the thermal pressure of the ambient medium is negligible.

As the jet velocity is typically close to the escape velocity, $v_j\propto (M/R)^{1/2}$ (where $M$ and $R$ are the mass and radius of the object which is launching the jet), then equation \ref{eq2} leads to  $a_0\propto R/M$. Since for MS stars, $R \propto M^{0.8}$ \citep[e.g.][]{popper1980}, then $a_0$ depends weakly on the mass of the main sequence star ($a_0 \propto M^{-0.2}$). Thus, the results obtained in this paper, in particular where the jet is successful or choked, can be generalized to low mass stars ($M \lesssim 1$ $M_\odot$). A very different outcome is expected in the case of neutron stars and black holes in which the radius is several orders of magnitude smaller than that for MS stars, thus, $a_0$ is much smaller and jets are energetic enough to propagate successfully at any orbital separation \citep{mm2017}. The accretion rate in all stages was computed by: i) considering the enclosed mass in the orbital separation using the initial density profile (and assuming spherical symmetry to calculate $v_{kep}$), and ii) using the density from the original profile close to the location of the MS star at each stage. The accretion rate found in the grazing stage is $\sim 3 \times 10^{23}$~g~s$^{-1}$, in the pib $\sim 2 \times 10^{24}$~g~s$^{-1}$, and in the piww $\sim 10^{25}$~g~s$^{-1}$. The accretion rates we find in all stages are approximately 1-10\% of the correspondent BHL accretion rate. The latter is consistent with the mass accretion rate onto a neutron star (or a black hole) immersed within the CE of a massive star \citep{macleod2015, mm2017, gilkis2019, lc2019, lc2020}. Thence, the successful jets during the grazing and beginning of the {\it{plunge-in}} stage require luminosities of order $L_j \gtrsim 10^{37}$~erg~s$^{-1}$ and $L_j \gtrsim 10^{39}$~erg~s$^{-1}$, respectively. When the MS star is well within the CE, according to equation 2, the jets are choked even for $10^{40}$~erg~s$^{-1}$. Also, the successful self-regulated jet models need less power than the constantly powered models (by a factor of $\sim$2-3). For the jets to be successful they need to clear the region close to the injection boundary. This is done more efficiently if the jets self-regulate and their luminosity adjusts to the changing conditions of the accreting environment. 

The Eddington luminosity is $L_{\rm{Edd}} \sim 10^{38} (M/M_{\odot})$~erg~s$^{-1}$, while the release of the binding energy of the accreting material onto the stellar surface is five to a hundred times larger. Thus, we have super-Eddington luminosity powering the jets. We notice that to properly determine the fate of the material crossing the inner boundary requires simulations which include radiation transport (due to the super-Eddington accretion) and in which the inner boundary is set at the stellar surface of the MS star (which is a factor of $\sim$ 3 smaller than our inner boundary).

We have assumed that the jets from each model are emitted at a stage of the MS and RG system evolution (either during the grazing-stage, or at the commencement or termination of the plunge-in stage). The delay in the launching of the jets may be due to the fact that an accretion disk around the MS star may not have formed, or that the jet launching mechanism comes into play. The disk can be formed through two main channels, either by the accretion of the CE material, or if the disk was pre-existent (i.e., formed before the MS star was engulfed by the RG and survived the grazing and plunge-in phase). The critical specific angular momentum to form a centrifugally supported disk at the inner boundary is $J_{\rm kep, in} = R_{\rm in} v_{\rm kep, in}$, with $v_{\rm kep, in}$ the Keplerian velocity at the inner boundary ($\approx 2\times 10^{7}$~cm~s$^{-1}$). Thus, $J_{\rm kep, in} = R_{\rm in} v_{\rm kep, in} \approx 2\times 10^{18}$~cm$^{2}$~s$^{-1}$. Since the velocity in the orbital plane at $R_{\rm in}$ is $\approx 10^{7}$~cm~s$^{-1}$, then the specific angular momentum at the inner boundary is $J_{\rm sp, in} \approx 10^{18}$~cm$^{2}$~s$^{-1}$. The latter corresponds to $\sim$0.5$J_{\rm kep, in}$, thus, our calculations show that it is unlikely that all accreted material will be accreted through a disk but a fraction of it possibly will. Our simulations lack magnetic fields and other angular momentum transfer mechanisms (for example tidal synchronization, shear, viscosity, etc, see \citet[][]{armitage2022} for further details), thus our angular momentum values should be considered as un upper limit.

The MS star is expected to enter the common envelope with a disk from a pre-phase of RLOF. Such disks are observed in several common envelope simulations, e.g., S19,  \citet{staff2016}, \citet{reichardt2019}. The question is whether such disks extend to the smaller scale around the companion and whether accretion can take place at such disks. For this purpose, simulations that include accretion modeling and that resolve the companion need to be carried-out. In our simulations, using a sub-grid accretion model with which we are close to resolving the MS star we find that the critical angular momentum to form a centrifugally supported accretion disk at the surface of a $0.3M_\odot$ star ($R_{\rm surf} \sim 3 \times 10^{10}$~cm) is $J_{\rm kep, surf}  \simeq 10^{18}$~cm$^{2}$~s$^{-1}$. If the angular momentum is conserved between $R_{\rm in}$ and $R_{\rm surf}$, then our simulations indicate that a small centrifugally supported disk may form close to the surface of the MS star (since the specific angular momentum may be larger than $J_{\rm kep, surf}$). In order to properly address the disk creation around the MS simulations with a smaller $R_{\rm in}$ than that in our study, simulations which include radiation transport and magnetic fields \citep[which properly transport the angular momentum, ][]{balbushawley1991} are required. Given that binary systems that undergo a CE phase have to previously experience a Roche-lobe overflow (RLOF) mass-transfer phase, thus, it is likely that a  accretion disk may have formed around the MS star before it was engulfed by the CE. The latter has been observed in CE simulations that start with a companion far from the RG ($a \geq 2 R_{\rm RG}$) and in which a disk is formed around the MS star \citep[in particular when the pre-RLOF stage is simulated][]{shiber2021}. It is also possible that the sub-Keplerian inflows will form an accretion belt around the accreting object (instead of an accretion disk) from which jets may be launched \citep{schreier2016}, or a polar outflow may be produced without the presence of an accretion disk due to an inflow with an equatorial-to-polar density gradient \citep{lery2002, aguayo2019, aguayo2021}. 

It is unclear what would be the long-term evolution of the system in the case of choked jets. The continuous injection of energy from the star/disk system into the jets will be deposited in the region close to the star. If the ram pressure of the accreted material is much larger than the jet ram pressure, the jet material flow will be reversed and it will go back to the disk, forming a small ``circuit'' of ejected-accreted material (in which most likely the accreted jet material will be a fraction of the accreted envelope material). Detailed numerical simulations are required to compute this precisely.

The setups used as the initial condition of our simulations (at the grazing, pib and piww stages) taken from S19 did not consider jets. Nevertheless, the difference in the orbital separation and in the mass ejection when considering jets is negligible (though the geometrical envelope structure is somewhat different). Thus, our results can be compared with the large scale models of S19 that do include jets. Both studies agree on the fact that, in the grazing stage, even jets with low $\eta$-efficiencies may be successful. Also the jets are diverted outwards by the envelope of the RG, thus, the resulting geometrical structure produced around the edge of the envelope and the MS star is very similar. At the pib stage, the jets of the large scale simulation successfully propagate outside, while in the local simulation they are only successful for relatively large $\eta$ values. In the piww stage the jets are initially choked inside the  envelope in both simulations. However, in the large scale simulation the jets are able to drill outside on a time scale of 3 days, whereas in the small scale simulations they are still choked after $\sim10$ days.

Based on the luminosity estimates of order $L_j \gtrsim 10^{37-39}$~erg~s$^{-1}$ and integrating over jets energy deposition timescales of order $10^2-10^3$days (as obtained in the 3D HD large scale simulations of S19), we have total energy deposition between $E_j \gtrsim 10^{44-47}$~ergs. Considering that the binding energy of the RG star is $E_{\rm bind} \simeq 10^{47-48}$~ergs, hence, most of the CE will remain bound. However, since the binding energy of the outer part of the envelope of the RG (i.e. that for $R>20 R_{\rm \odot}$) is $\sim 10^{46}\; {\rm ergs}$, thus, jets may facilitate partial CE-envelope removal (including other sources of energy such as recombination, magnetic fields, enthalpy could help to fully remove the CE, see \citet{ivanova2013} for further details). A consequence of the partial CE-stripping would be that less material is present during the plunge-in stage, and the jets may be successful. If the CE of the RG is mostly convective, which would be the case when the low mass RG has evolved from its respective terminal main sequence stage, then the material is less bound \citep{shorelivio1994,klencki2021}. Thus, the CE would be lifted more efficiently by the jets.

The total amount of accreted mass in our simulations is still relatively high. On the short time scale we simulated ($10^6\;{\rm s} \approx 11.6$~days), the MS star accretes $\sim 10^{-2} M_{\odot}$ while plunging-in. This is $\sim$1\% of the total mass we had initially in the grid during the correspondent stage. If such an accretion rate persists for a typical duration of $\sim 500$~days in which the envelope has not been removed yet, the MS star can accrete $\sim 0.5 M_{\odot}$. Assuming a similar amount of accretion in a CE event while the primary is an AGB star, it is possible that the carbon enrichment of dwarf carbon stars, which have been recently observed in short-period binary systems of less than a day \citep{roulston2021, whitehouse2021}, can take place also during the CE.

The photospheric temperature of the CE is $T_{\rm eff} \sim 10^3$~K, thus, its emission properties will likely resemble those of a RG and be observable in the NIR-optical\citep[see for example:][]{valenti2004, huang2020}. As the medium is optically thick, propagation of the jets through the CE may produce a bright quasi-thermal spectrum. Jets powered by high mass accretion during the GEE could produce a transient event since a non-negligible fraction of the kinetic energy of the jets may be be transferred to radiation energy \citep{soker2016}. While a full radiation HD calculation is needed to properly compute the observational outcomes of the jet-cocoon system breaking out of the CE, as well as those of the ejected stellar material, bright thermal emission is expected. From the Rankine-Hugoniot jump conditions the temperature of the post-shock material can be estimated by $kT = 3/16 m_p v_{\rm sh}^2$, where $v_{\rm sh}$ is the shock velocity. Thus, for shock velocities or order $v_{\rm sh} \sim 10^7-4\times 10^7$~cm~s$^{-1}$ the temperature is $\sim 10^5 - 10^6$~K. As the region close to the MS star is opaque (as its optical depth is $\tau \sim \rho k R \gg 1$), the emission will be thermal and its luminosity\footnote{The large uncertainty in the luminosity is direct consequence in the uncertainty in the temperature in the photosphere (which is not spatially resolved in our simulations).} will be of order $L=4\pi \sigma R^2 T^4 \sim (10^{39}-10^{43}) \ R_{11}^2$~erg~s$^{-1}$. Thus, the spectrum will peak in X-rays and UV at early times and evolves to lower frequencies as a function of time. Transients with these luminosities are visible up to distances of $\approx$ 1-100 Mpc.

The jets injection rate of simulation \#6 in S19 was 0.001 M$_{\odot}$~yr$^{-1}$, which was 0.66\% of the BHL accretion rate calculated at the RG surface $\sim$0.15 M$_{\odot}$~yr${^-1}$. However, we find much lower accretion rates in the grazing (30 times smaller) and pib (5 times smaller) stages, and about the same accretion rates in the piww stage. This would imply efficiency values of 20\%, 3.5\%, and 0.7\% of the jets in S19, at each stage, respectively. Therefore, it might be that the jets power in S19 is overestimated in the grazing and pib phases, although these efficiency values are still reasonable considering the fact that power of the jets in S19 was constant through the simulation evolution. As the main effect of the jets on the orbital evolution and on the envelope, unbinding starts only afterwards (t>53~days), this overestimation most likely did not play a significant role.

On the other hand, because most of the unbinding in S19 happens afterwards when the MS is deep inside the envelope, the question whether the jets are successful to penetrate through the envelope or choked at this stage is critical. In our simulation the jets are choked well within the envelope and therefore are not expected to be able to unbind the envelope. There are two factors which make a difference on whether the jets propagate or not in the simulations of S19 and the ones in this paper: 1) The injection radius in our case is 10$^{11}$~cm, while in S19 is 10$^{12}$~cm. Thus, our jets have to overcome a larger amount of ambient medium with a higher ram pressure. The energy necessary to move the extra mass, from deeper within the gravitational potential, is not considered in simulations that launch the jets further outside. 2) The boundary conditions responsible for the jet injection in S19 are different with respect to our paper. In both simulations the jets are injected from a cone with an opening angle of 30$^\circ$. While in S19 the density is redefined at each timestep, in our simulations both density and pressure are redefined at each timestep. To better address this question a future study that explores in detail the numerical parameters involved in the simulations is required.

\section{Conclusions}
\label{sec:conc}
In this paper, we have presented small scale 3D numerical simulations of the propagation of jets through a CE formed by a 0.88 $M_\odot$ red giant and a 0.3 $M_\odot$ main sequence star. The simulations were based on the CE structure configuration resulting from the large scale simulations of \citet{shiber2019}. We find that the final outcome depends on the conditions of the environment, on the power of the jet, and whether the jet is constantly powered or self-regulated. In the grazing stage and the commencement of the plunge self-regulated jets need higher efficiencies to break out of the envelope of the RG. Deep inside the CE, on the timescales simulated, jets are choked independently of whether they are self-regulated or constantly powered. Jets with the same characteristics as those in the large-scale simulations of S19 (which are able to break out of the envelope of the RG), are choked in our small-scale simulations (for at least the integration time, $\sim$11~days). Having a smaller jet injection radius, and the self-regulation of the jets may affect the outcome. Our simulations are a step in the direction of investigating the interface between local and global simulations.

Once jets are launched, their success and propagation through the CE depends on the conditions of the environment as well as on the power of the jets. Self-regulated jets will be by their nature more efficient than the constantly powered case. These are more luminous when the accretion rate is high, and less luminous when the accretion rate is small. On the other hand, constantly powered jets are always using the same energy deposition rate, regardless of the accretion rate. Also, high luminosity emission is expected in this case going from X-rays to UV and optical as a function of time. Deep inside the CE the jets are choked since the ram pressure of the jets is overcome by that of the accreting material.

Our simulations show that the mass accretion onto the MS star is 1-10\% of the Bondi Hoyle Littleton rate ($\sim 10^{-3}-10^{-1}$ M$_\odot$~yr$^{-1}$), and that a disk around the secondary MS star is probably not formed. Still, jets could be launched in the case of a pre-existing disk or via MHD effects. It is unclear if the disk can survive when it is engulfed into the CE and needs to be further studied. Future surveys (including the Vera C. Rubin Observatory \footnote{www.lsst.org}) will help to decipher the CE phase. The outlook for the study of the CE phase is promising as it is likely that many of the observed HEAP events will allow to place constraints on the models. 

We must note that we only follow $\sim$11 days in each of the three stages, thus we do not follow the totality of the interaction of jets within the CE. Future large scale simulations with high-resolution, which are currently at the edge of our computational capabilities, will permit to fully study the evolution of jets, as well as the disk creation, when immersed in the CE of a massive star.

\section*{Acknowledgments}
We thank the referee for the helpful comments and improvement of the manuscript. D.L.C. is supported by C\'atedras CONACyT at the Instituto de Astronom\'ia, UNAM. We acknowledge the support from the Miztli-UNAM supercomputer (projects LANCAD-UNAM-DGTIC-321 and LANCAD-UNAM-DGTIC-281) for the assigned computational time in which the simulations were performed. D.L.C. and F.D.C. thank the UNAM-PAPIIT grant IG100820. S.S. thanks National Science Foundation Award 1814967 for its support. R.I. is grateful for the financial support provided by the Postodoctoral Research Fellowship of the Japan Society for the Promotion of Science (JSPS P18753). Many of the images in this study were produced using VisIt. VisIt is supported by the Department of Energy with funding from the Advanced Simulation and Computing Program, the Scientific Discovery through Advanced Computing Program, and the Exascale Computing Project.

\section*{Data availability}
The data underlying this article will be shared on reasonable request to the corresponding author.

\bibliographystyle{mn2e}
\bibliography{bibliography}

\end{document}